\documentclass[reprint,showpacs,preprintnumbers,amsmath,amssymb,aip,nolongbibliography]{revtex4-1}

\usepackage{amssymb}
\usepackage{graphicx}
\newcommand{\tx}{\tau_{\mathrm{x}}}
\newcommand{\tp}{\tau_{\mathrm{p}}}

\begin{document}

\preprint{1}

\title{Characterizing heterogeneous dynamics at hydrated electrode surfaces}

\author{Adam P. Willard}
\affiliation{Department of Chemistry and Biochemistry, University of Texas, Austin, Texas 78712}
\author{ David T. Limmer}
\affiliation{Department of Chemistry, University of California, Berkeley CA, 94720}
\author{Paul A. Madden}
\affiliation{Department of Materials Science, Oxford University, Oxford OX1 3PH, United Kingdom}
\author{David Chandler}\email[Corresponding author.  E-mail: ]{chandler@berkeley.edu}
\affiliation{Department of Chemistry, University of California, Berkeley CA, 94720}

\begin{abstract}
In models of Pt 111 and Pt 100 surfaces in water, motions of molecules in the first hydration layer are spatially and temporally correlated.  To interpret these collective motions, we apply quantitative measures of dynamic heterogeneity that are standard tools for considering glassy systems.  Specifically, we carry out an analysis in terms of mobility fields and distributions of persistence times and exchange times.  In so doing, we show that dynamics in these systems is facilitated by transient disorder in frustrated two-dimensional hydrogen bonding networks.  The frustration is the result of unfavorable geometry imposed by strong metal-water bonding.  The geometry depends upon the structure of the underlying metal surface.  Dynamic heterogeneity of water on the Pt 111 surface is therefore qualitatively different than that for water on the Pt 100 surface.  In both cases, statistics of this adlayer dynamic heterogeneity responds asymmetrically to applied voltage.
\end{abstract}

\maketitle
The behavior of water adjacent to hydrated metal surfaces is central to the physics underlying electrochemistry, and as such there is much theoretical work devoted to this topic.  References~\onlinecite{GAV01,AM01,SWG02,AM03,AM04,SM05,YC06,AM07,AM07b,SM07,YC07, JIS95, APW09, DL12} are examples.  Our recent work in this area~\cite{DL12} highlighted the role of interfacial properties occurring over nanosecond timescales.  In this paper, we examine this dynamics in further detail, considering models of Pt-water interfaces and showing structural relaxations in the water ad-layers proceed exceedingly slowly via mechanisms that are spatially heterogeneous.  

Strong metal-water bonding forces ad-layer waters into structures that are antithetical to favorable hydrogen bonding.  Accordingly, the ad-layers contain defects.  Motions are one- to two-orders of magnitude faster in the proximity of these defects than elsewhere on the surface.  Further, rearrangements of these defects require coordinated motions of several water molecules.  This type of heterogeneous dynamics is common in glass-forming materials.\cite{MDE96,CG09}  Some of the quantitative methods used to successfully interpret such behavior in molecular dynamics models of glass formers\cite{CG09,KHGC11} are used here to elucidate the nature of water dynamics at metal surfaces.

 \begin{figure*}[t]
\includegraphics[width=7in]{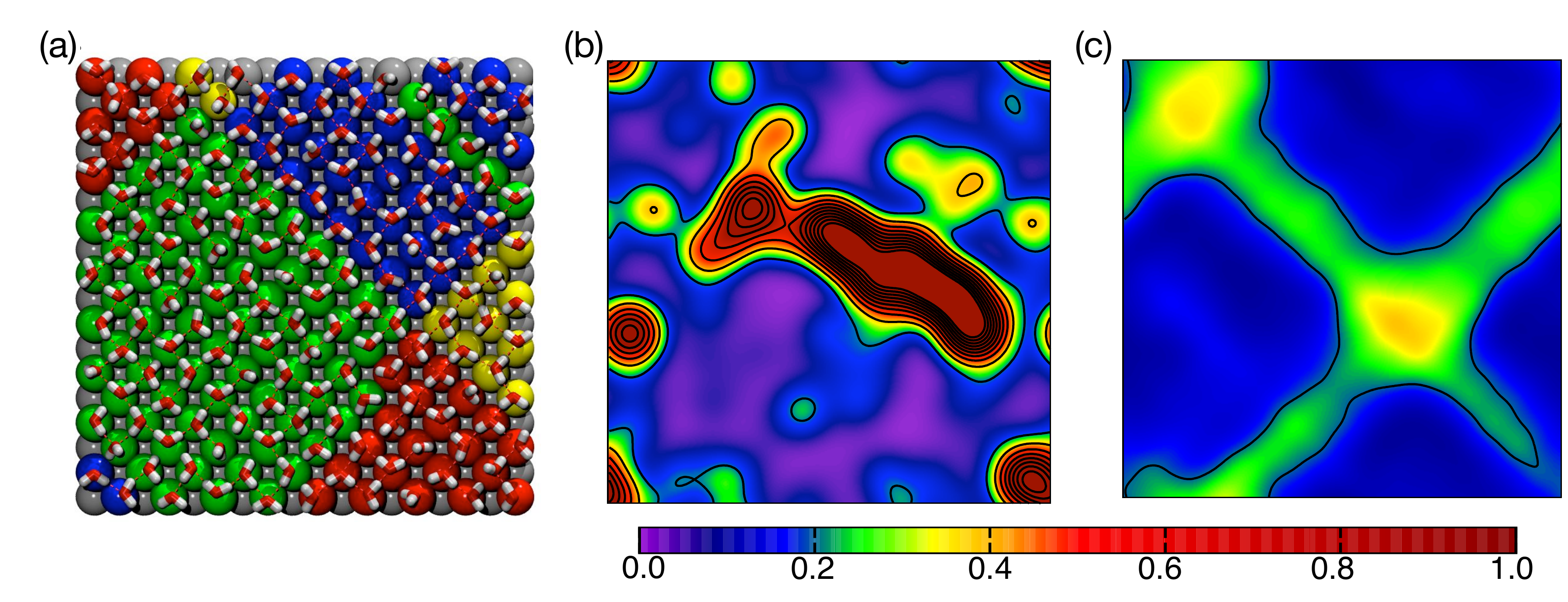}
\caption{An instantaneous configuration and dynamic heterogeneity of the water ad-layer on the Pt 100 surface.  The ad-layer is in equilibrium with adjacent bulk water (not shown). (a) Hydrogen bonding patterns showing heterogeneous distribution of line defects.  Colors of underlying electrode atoms highlight distinct domains of ad-layer waters with specific hydrogen bond arrangements.  (b) and (c) Instantaneous and time averaged mobility fields, $q(\mathbf{a};t)$ and $\bar{q}(\mathbf{a};t_\mathrm{obs})$, respectively, with $t_\mathrm{obs} = \tp/3$. The snap shots in (a) and (b) are taken at the midpoint of the trajectory that is averaged to produce (c).  Color code for the mobility field is given by $q(\mathbf{a},t) \Delta x^3$ where $ \Delta x = 0.1 \mathrm{\AA}$.}
\label{fig:qifield100}
\end{figure*}

\subsection* {Model}
As in our earlier work,\cite{APW09, DL12} we carry out molecular dynamics simulations with a surface model of Siepman and Sprik.\cite{JIS95}  The model provides an empirical water-metal chemisorption potential. It includes effects of the metal's electronic polarizability and its response to applied voltage across an electrochemical cell.  For the model parameters we consider, the electrodes most resemble platinum surfaces\cite{HH09, MRA03,TM87, FKSOZK11} with water-metal binding energies around $0.4 e\mathrm{V}$ and a metal lattice constant of 4.0\AA. Consequently when the electrode is exposed to a bulk liquid reservoir, water molecules adsorb to the metal atoms and adopt orientations that allow for hydrogen bond formation with molecules adsorbed to adjacent lattice sites. Snapshots of the ad-layers formed spontaneously from a slab of liquid water are shown in Panels (a) of 
Figs.~\ref{fig:qifield111} and~\ref{fig:qifield100}, which refer to Pt 111 and Pt 100 surfaces, respectively.  The snap shots show configurations of the ad-layer waters only.  The liquid in contact with the ad-layer is not shown for purposes of clarity.  Other pictures of this system can be found in Refs.~\onlinecite{APW09, DL12}. 

Depending on the surface geometry, adsorbed water molecules form qualitatively different hydrogen bonding patterns.
For the 100 surface, the four-fold coordination and lattice spacing are commensurate with a variety of two-dimensional hydrogen bonding patterns. Surface water molecules adsorb to nearly every available surface site and vacant surface atoms are rare. The surface patterns that emerge are domains distinguished by relative orientations of the adsorbed water molecules.  Different colors in Fig.~\ref{fig:qifield100}(a) help identify different domains. Transient disorder in the form of line defects between domains of dipole aligned molecules relax slowly, as we shall see. 

For the 111 surface, the six-fold lattice coordination is incommensurate with planar arrangements that would allow all molecules to form four hydrogen bonds. In this case, the particularly stable configurations of water molecules involve the formation of three hydrogen bonds where a given water molecule donates two and accepts a single hydrogen bond in a triangular manner with molecules residing on alternating coordination sites.  Blue coloring in Fig.~\ref{fig:qifield111}(a) helps identify these patterns.  These particular arrangements are facilitated by the presence of nearby surface vacancies that help eliminate hydrogen bond frustration and account for 15\% of all surface sites on average.  Since vacancy diffusion is slow, the spatial distribution of vacancies is heterogeneous over relatively long times, as shown below. 

\subsection*{Orientational mobility and fields}
The chemisorption energy (i.e., the metal-water bonding energy) is sufficiently strong that lateral diffusion of water is rare on picosecond timescales. As such, the dynamics of an adsorbed water molecule are dominated by rotations. To characterize changes in molecular orientations, we focus on the variables $\mathbf{u}_i(t)$, which denote the unit vector parallel to the dipole of the $i$th water molecule at time $t$.  

There are different contributions to the time dependence of $\mathbf{u}_i(t)$ that act on different time scales.  In typical trajectories, $\mathbf{u}_i(t)$ oscillates rapidly about a reference direction before jumping to new reference direction.  The oscillations reflect small-amplitude vibrations and librations of the water molecule, and the jumps (or instantonic events) reflect changes in long lived arrangements of several molecules. The former generally have periods of 1 ps or less, while distinct molecular arrangements generally persist for more than 5 ps, and usually much longer.  Indeed, as we detail later, the mean persistence time is 20 ns and 1 ns for the water ad-layers on the Pt 100 and Pt 111 surfaces, respectively.

As a result of this separation of time scales, it is useful to consider
\begin{equation}
\bar{\mathbf{u}}_i(t) = \frac{1}{\delta t} \int_t^{t+\delta t} d t' \mathbf{u}_i(t').
\end{equation}
The value of the coarse-graining time, $\delta t$, should be large enough to remove most vibrational and librational contributions to $\bar{\mathbf{u}}_i(t)$, and thereby highlight contributions from structural reorganization.  For this purpose, we use $\delta t = 2$ ps.  Distribution functions graphed later in this paper show that this value is indeed suitable. 

In terms of these coarse-grained coordinates, an instantaneous measure of mobility is given by
\begin{equation}
q_i(t) = 1-\bar{\mathbf{u}}_i(t) \cdot \bar{\mathbf{u}}_i(t+\Delta t).
\label{eq:q}
\end{equation}
To the extent that it is non-zero, the configuration at time $t+\Delta t$ differs from that at time $t$. To use this quantity to count relevant reorganization events, $\Delta t$ must be both greater than the coarse-graining time, $\delta t$, and much less than the time to de-correlate enduring orientations. For this purpose, we use $\Delta t=10\,\mathrm{ps}$.  This choice is suitable for describing ad-layer mobility because 10 ps is one- to two-orders of magnitude shorter than the relevant mean persistence times.
 \begin{figure*}[t]
\includegraphics[width=7in]{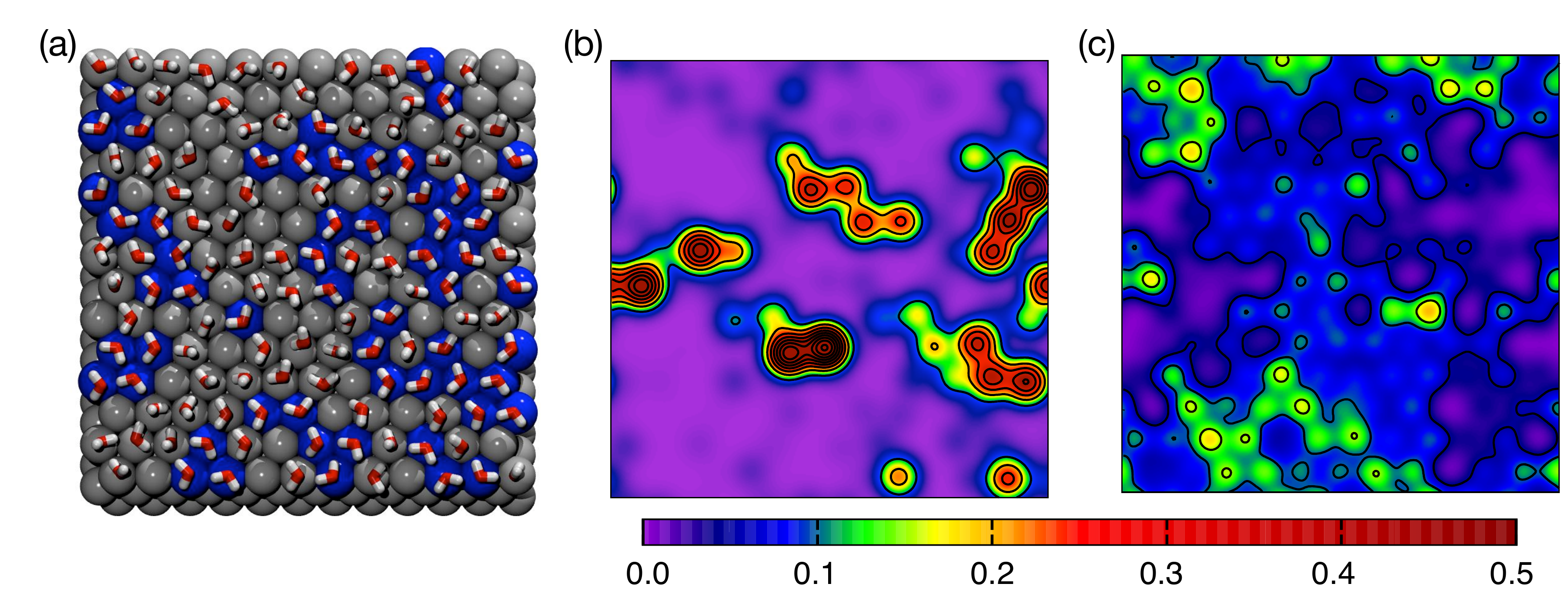}
\caption{An instantaneous configuration and dynamic heterogeneity of the water ad-layer on the Pt 111 surface.  The ad-layer is in equilibrium with adjacent bulk water (not shown). (a) Hydrogen bonding patterns: water molecules engaging in the preferred hydrogen bond pattern (see right-hand side of panel) have their underlying electrode atom colored blue. (b) and (c) Instantaneous and time averaged mobility fields, $q(\mathbf{a};t)$ and $\bar{q}(\mathbf{a};t_\mathrm{obs})$, respectively, with $t_\mathrm{obs} = \tp/3$. The snap shots in (a) and (b) are taken at the midpoint of the trajectory that is averaged to produce (c).  Color code for the mobility field is given by $q(\mathbf{a},t) \Delta x^3$ where $ \Delta x = 0.1 \mathrm{\AA}$.}
\label{fig:qifield111}
\end{figure*}
Spatial resolution of mobility can be resolved with reference to a mobility field.  Our specific choice mobility field focuses on ad-layer molecules and is coarse grained over a length $\xi$.  In particular, we define
\begin{equation}
q(\mathbf{a},t) =  {\sum_i} q_i(t) \phi(\mathbf{a} - \mathbf{a}_i(t); \xi) \Theta(2\xi - |z_i(t)-z^*|).
\end{equation}
Here, $\mathbf{a}_i(t)$ is the two-dimensional projection of the $i$th water-oxygen position onto the plane of the electrode, and $z_i(t)$ is the projection perpendicular to the plane; $z^*$ is the mean of $z_i(t)$ for water molecules in the ad-layer; $\Theta (x) = 1$ or $0$ for $x \geqslant 0$ or $x<0$, respectively; $\phi(\mathbf{a}; \xi)$ is a delta-like function broadened over length scale $\xi$; we use a normalized Gaussian, $\phi(\mathbf{a}; \xi) \propto \exp(-a^2/2\xi^2))$, and set $\xi=1.5 \mathrm{\AA}$. 

Snapshots of $q(\mathbf{a},t)$ for the 111 and 100 surfaces are shown in 
Figs.~\ref{fig:qifield111}(b) and \ref{fig:qifield100} (b), 
respectively. Both surfaces display significant heterogeneous dynamics, manifested as regions of high mobility within a background of low mobility. For the 100 surface, the high mobility regions have a definite directionality in that they are aligned with the underlying lattice, and they are diffuse in that they encompass several neighboring molecules. Reorganization is  therefore collective, occurring along existing hydrogen bond chains.  Further, by transiently disrupting the local hydrogen bond network, reorganization of one molecule facilitates reorganization of neighboring molecules.  Such behavior is characteristic of glassy systems in general.\cite{CG09}  

For the 111 surface, the high-mobility patches also have a directionality, though the underlying lattice dictates that motion does not propagate in a straight line but rather jigsaws back and forth along the lattice. Mobile regions are more localized on the 111 surface compared to the 100 surface, but motion is similarly facilitated as evidenced by the connectivity of the mobile regions.  Movies illustrating the time-evolution of dynamic heterogeneity for the water monolayers on the 100 and 111 surfaces can be viewed at http://youtu.be/cCYLPtckDWk and http://youtu.be/d0DDWuP-qTg, respectfully.

The time scales over which immobile and mobile regions interconvert are the time scales of persistence.  The mean persistence time, $\tp$, is the structural relaxation time.\cite{CG09} Over times smaller than $\tp$, patches of high mobility occur with higher frequency at particular regions of space. The persistent structural degrees of freedom contributing to these motions can be highlighted by time-averaging the spatial mobility field,
\begin{equation}
\bar{q}(\mathbf{a};t_\mathrm{obs}) = \frac{1}{t_\mathrm{obs}} \int_0^{t_\mathrm{obs}} dt \,q(\mathbf{a},t).
\end{equation}   
where $t_\mathrm{obs}$ is an observation time. The time averaging causes mobile domains to become diffuse, as the higher frequency features disappear.  Heterogeneity disappears completely as 
$t_\mathrm{obs}/ \tp$ becomes large.

For the 100 surface, this time averaging elucidates the role of line defects on the surface as promoters of reorganization. Shown in Fig.~\ref{fig:qifield100}(c) for $t_\mathrm{obs}=\tp/3$, integrated motion is primarily along lines that are fully connected and span the entire system. The linear structures run parallel to the surface close packing, and orthogonal to each other. The features can be readily identified with persisting line defects on the surface by analyzing the hydrogen bonding patterns. Figure ~\ref{fig:qifield100}(a) shows the molecular configuration at the midpoint of the trajectory used to calculate Fig.~\ref{fig:qifield100}(c), where the underlying metal lattice has been colored for one of four hydrogen bonding patterns (also shown). The location of line defects formed at the boundaries between these different domains coincides with the regions high mobility.

For the 111 surface, time averaging the mobility field yields similar results. As shown in Fig.~\ref{fig:qifield111}(c)  for $t_\mathrm{obs}=\tp/3$, integrated motion exhibits domains of high mobility. These domains, like the ones on the 100 surface, become more connected although their structure is not as simple. The locations of the domains are not directly related to the location of surface vacancies but rather are spatially anti-correlated to the presence of stable hydrogen bond configurations.  Molecules with this preferred hydrogen bonding pattern typically reside within inactive patches.  Figure~\ref{fig:qifield111}(a) is taken from the midpoint of the trajectory used for Fig.~\ref{fig:qifield111}(c) and designates these particularly stable hydrogen bond configurations as blue lattice sites. 

\begin{figure}[b]
\includegraphics[width=8.5cm]{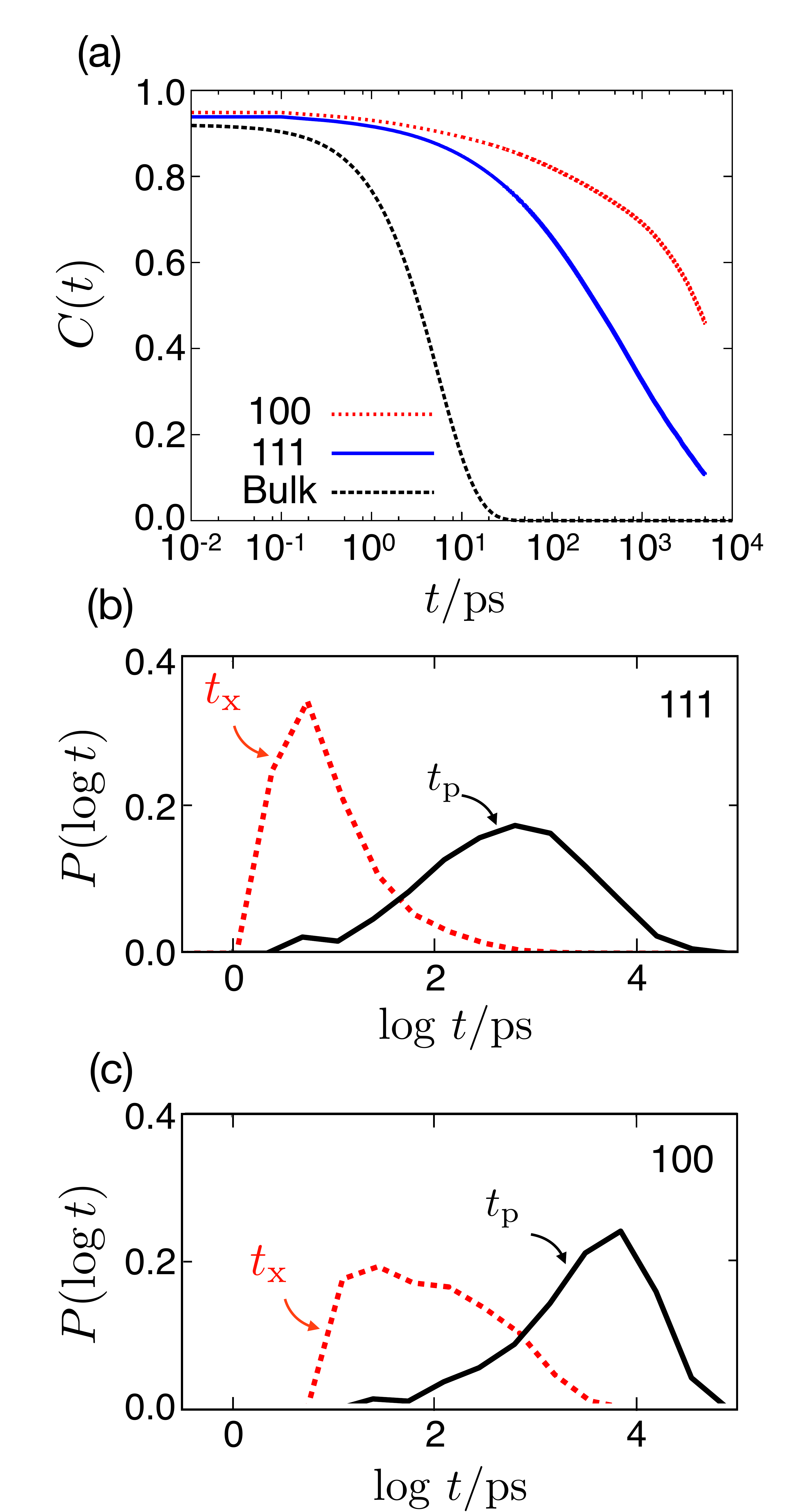}
\caption{(a) The dipole autocorrelation function for water molecules adsorbed to the 100 (dotted red line) and 111 (solid blue line) electrode. The corresponding quantity for molecules in the bulk liquid is plotted as a dashed black line. (b) The probability distributions for the persistence, $t_\mathrm{p}$, and exchange times, $t_\mathrm{x}$, for the 111 surface, shown on a log scale where $P(\log\,t)=t P(t)$. (c) The probability distributions for the persistence and exchange times for the 100 surface.}
\label{fig:tcorr}
\end{figure}

\subsection*{Correlation and distribution functions}

These heterogeneous, facilitated dynamics described above dictate that the timescales associated with relaxing a given tagged molecule will be large relative to the bulk liquid, and strongly dependent on its environment. The average timescale governing orientational reorganization can be extracted from the dipole autocorrelation function,
\begin{equation}
C(t) = \langle \mathbf{u}(0) \cdot \mathbf{u}(t) \rangle - |\langle \mathbf{u} \rangle|^2,
\label{eq:tcorr}
\end{equation}
where the angle brackets represent an average over all electrode-adsorbed water molecules. Note that while $\langle \mathbf{u} \rangle$ vanishes in the bulk liquid, it does not necessarily at the surface. As shown in Fig.~\ref{fig:tcorr}(a), water on both surfaces undergo orientational relaxation over timescales that are of 2 to 20 ns. The more ordered 100 surface yields significantly slower relaxation than for that of the locally over-coordinated 111 surface. These times are 3-4 orders of magnitude larger than the characteristic relaxation time for bulk water, $\tau_\mathrm{bulk}=5~\mathrm{ps}$.  (For the computation of $C(t)$ for bulk water, the averaging implied by the angle brackets is carried out over all non-ad-layer waters.) 

The functional form of the decay of $C(t)$ for bulk water is an expected exponential decay, indicative of an uncorrelated Poisson process characterized by a single timescale, $\tau_\mathrm{bulk}$. The functional form for the decay of $C(t)$ for the adsorbed water, however, is more complicated and resembles the stretched exponential decay associated with correlation functions of glass-forming liquid systems.\cite{MDE96} Such deviations from Poisson statistics highlight the important temporal correlations within the orientational dynamics. 

These temporal correlations are manifested explicitly in the distributions of persistence and exchange times.\cite{CG09}  Persistence times, $t_\mathrm{p}$, are the waiting times for reorientations, and exchange times, $t_\mathrm{x}$, are the times between such motions. In cases of uncorrelated dynamics, persistence and exchange times are drawn from the same distribution, fully characterized by the decay time of $C(t)$. However, if the dynamics are temporally correlated then the probability for undergoing motion is conditionally dependent on what has already occurred, and these distributions will be therefore different.\cite{GG01}  In glassy dynamics, where motion in a region of space facilitates further motion in neighboring regions of space and time, the most probable $t_\mathrm{x}$ is much smaller than the most probable $t_\mathrm{p}$.

This decoupling of exchange and persistence is clearly evident in Figures \ref{fig:tcorr}(b) and (c), which depict distributions of persistence and exchange times for both the 111 (b) and 100 (c) surfaces. The times are calculated by recording the time for a dipole to change its orientation by $50^{o}$ either from an arbitrary time origin, for the persistence time, or given a reorientation just occurred, for the exchange time. This angular displacement is chosen to be much greater than the librational motion of water on the surface and commensurate with reorientations between stable hydrogen bonding patters on the surface, as dictated by the lattice geometry. In order to avoid counting reorganizations that only occur transiently, we also stipulate that the rotation persists for an additional $\Delta t$.  (For reference to similar calculations of exchange and persistence distribution, but for a model of a structural glass-forming liquid, see Ref.~\onlinecite{HMCG07}.)

For ad-layers to both metal surfaces, the mean exchange time, $ \tx $, is shorter by over an order of magnitude from the mean persistence time. The mean persistence time, $\tp$, agrees with the $1/e$ time from the decay of $C(t)$. This separation of timescales, reflecting a dynamic facilitation mechanism, occurs in here because transient disorder within the adsorbed monolayer gives rise to spatial variations in the relative hydrogen bond stability, which in turn facilitates heterogeneous dynamics. 

\begin{figure}[t]
\includegraphics[width=8.5 cm]{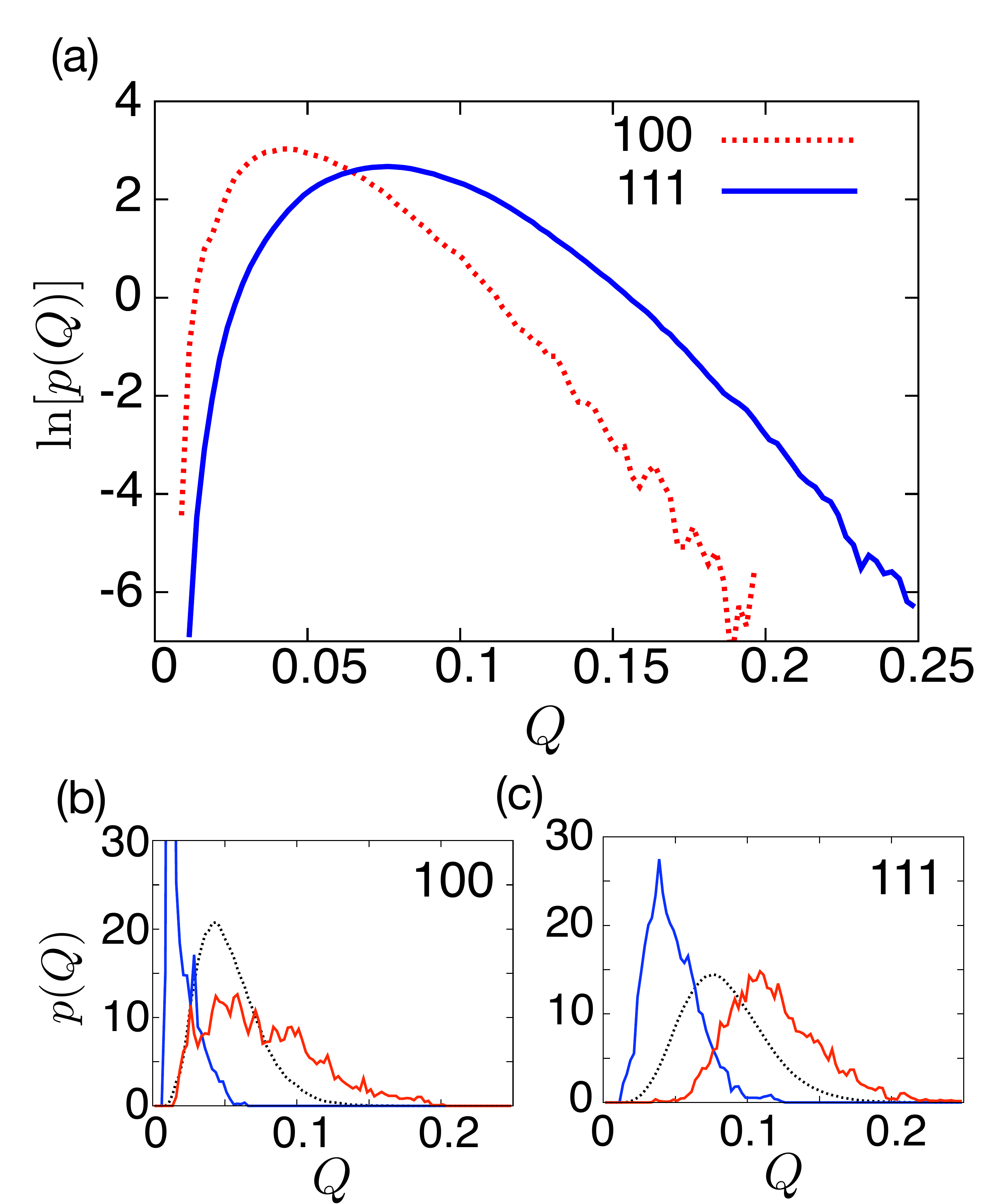}
\caption{(a) The probability distribution for the total orientational mobility, $Q$, plotted for the 100 electrode surface (dashed red line) and the 111 electrode surface (solid blue line). (b-c) The probability distribution $p(Q)$ at different values of the applied electrode potential. Dotted black line correspond to the results at zero applied potential.  Red and blue lines correspond to the results at the negative electrode ($V_0=-1.36\mathrm{V}$) and positive electrode ($V_0 = 1.36 \mathrm{V}$) respectively.}
\label{fig:pot}
\end{figure}

The total orientational mobility within the adsorbed monolayer is given by 
\begin{equation}
Q(t) = \frac{1}{N} \sum_{i=1}^N q_i(t) \, ,
\end{equation}
where the sum is carried out over all $N$ electrode-adsorbed molecules. ($N$ can fluctuate due to exchange between the bulk and ad-layer, but the resulting time dependence of $N$ has a negligible effect on $Q(t)$ for the conditions we have studied.)  This total orientational mobility variable serves as an order parameter distinguishing ad-layers of different mobility.  Figure~\ref{fig:pot} shows the distribution functions of of this order parameter, $p(Q)$ for the 100 and 111 electrode surfaces.  Both distributions are Gaussian around the mean, but both also possess fat tails at larger values of $Q$. The non-Gaussian tails indicate that highly active configurations are much more probable than one would expect given Gaussian statistics. The tails arise through the correlated nature of the surface dynamics.  Their presence indicates, in principal, that it is possible to drive the ad-layers out of equilibrium into distinct phases, one of high mobility and one of low mobility.\cite{GJLPW09}

This behavior does not change qualitativey under applied voltage.  Interestingly, the distributions of total mobility respond asymmetrically at the positive and negative electrode. Figure~\ref{fig:pot} shows $p(Q)$ for both electrode geometries and for the positive ($V_0 = 1.36\mathrm{V}$) and negative ($V_0 = -1.36\mathrm{V}$) constant potential electrode.  For both surfaces, the positive values of the applied potential have the effect of shifting the distribution of mobility to lower values of $Q$. This shift to lower $Q$ manifests as a slowing in the timescales for relaxation dynamics. At the negative potential electrode we observe the opposite effect, a shift in $p(Q)$ to larger values of $Q$, indicating an increase in the surface relaxation times. 

The asymmetric behavior can be understood in light of our previously published work in which we demonstrated that for moderate values of an applied electrode potential the orientations of adsorbed water molecules respond asymmetrically with respect to applied electrode potential.\cite{APW09}  The asymmetry arises because at the negative electrode adsorbed water molecules can adopt an alternative orientation in which one oxygen-hydrogen bond is pointed straight towards the electrode thereby gaining a favorable image charge interaction between the partial positive charge on the hydrogen and the electrode.  There is no corresponding orientation at the positive electrode and so at the negative electrode one relaxation pathway (rotation of an oxygen-hydrogen into the electrode) is enhanced while at the positive electrode the same pathway is inhibited.  This asymmetry was found to be more prominent for 111 versus 100 due to the relative instability of the hydrogen bond network over the 111.  

Consistent with the asymmetric behavior predicted from our model, others have noted that the electronic structures of metal interfaces respond sensitively to the electrostatic fields of the surrounding solvent.\cite{JA08,FKNV11} In particular, electronic structure theory\cite{JA08} indicates that dipole reorientations of adsorbed water molecules on a Pt 111 surface can result in shifts in the potential of zero charge of up to 3 $e$V. The heterogeneous relaxation elucidated herein are therefore expected to modulate catalytic activities of metal surfaces.  For a platinum-like surface, we expect modulating domains that extend over nanometers and that reorganize tens of nanoseconds.  It remains to be seen whether asymmetric response can be tapped as a means to drive electrodes out of equilibrium in a fashion that will produce a non-equilibrium transition between phases of high and low orientational mobility, and whether such a transition can be of practical use. 

\begin{acknowledgments} 
We are grateful to Aaron Keys for comments on an earlier version of the manuscript. Work on this project in its early stages was supported by the Helios Solar Energy Research Center of the U.S. Department of Energy under Contract No. DE-AC02-05CH11231. In its final stages, it was supported by the Director, Office of Science, Office of Basic Energy Sciences, Materials Sciences and Engineering Division and Chemical Sciences, Geosciences, and Biosciences Division under the same DOE contract number.
\end{acknowledgments}

\end{document}